\documentclass[twocolumn, amsmath,amssymb, superscriptaddress,nofootinbib]{revtex4-1}

\usepackage{amsmath,amsfonts,mathrsfs, amssymb}
\usepackage{epsfig,graphicx}
\usepackage{multirow,longtable}
\usepackage{array}
\usepackage{placeins}
\usepackage{color}
\usepackage[caption=false, font=footnotesize, justification=RaggedRight]{subfig}
\makeatletter
\g@addto@macro\appendix{\setcounter{figure}{0}}
\makeatother
\makeatletter
\g@addto@macro\appendix{\setcounter{table}{0}}
\makeatother
\graphicspath{{./Figs/}}

\begin{document}

\title{Author Impact Factor: tracking the dynamics of individual scientific impact}

\author{Raj Kumar Pan}
\author{Santo Fortunato}
\affiliation{Department of Biomedical Engineering and Computational Science, Aalto University School of Science, P.O.  Box 12200, FI-00076, Finland} 

\begin{abstract}
The impact factor (IF) of scientific journals has acquired a major role in the evaluations of the output of scholars, departments and whole institutions.  Typically papers appearing in journals with large values of the IF receive a high weight in such evaluations. However, at the end of the day one is interested in assessing the impact of individuals, rather than papers. Here we introduce Author Impact Factor (AIF), which is the extension of the IF to authors. The AIF of an author A in year {\it t} is the average number of citations given by papers published in year {\it t} to papers published by A in a period of $\Delta t$ years before year {\it t}. Due to its intrinsic dynamic character, AIF is capable to capture trends and variations of the impact of the scientific output of scholars in time, unlike the h-index, which is a growing measure taking into account the whole career path.
\end{abstract}

\maketitle
\section*{Introduction}

Nowadays, the impact of the work of a scientist is estimated by metrics. Typical metrics are the number of papers written by an author and the total number of citations received by these publications. 
Not all publication venues have equal prestige, though. It is valuable for a scholar to be able to publish, at least occasionally, on important journals/conferences. To compute the citation impact  
of journals, the {\it Impact Factor} (IF) was introduced~\cite{IF_essay}. The IF of a journal X at year $t$ is the average number of citations from papers published in year $t$ to papers of the journal X published in the two years preceding $t$ ($t-1$ and $t-2$). This measure, which is computed by Thomson Reuters~\cite{Thomson} every year for each journal of the Web of Knowledge database~\cite{WoK}, is currently used by academic and research institutions world-wide to weigh the importance of the output of papers and scholars. Scientists are ranked 
highly if they manage to publish some of their works in journals with large values of IF (e.g. {\it Nature}, {\it Science}, {\it Cell}, etc.). 

However, it is well known that the distribution of the number of 
citations of papers published in a journal is skewed, with most papers being poorly cited and a few being highly cited. For this reason the IF, which is an average over all published papers (in a given period), 
cannot depict well the impact of any of the papers, since the average of a broad distribution is not representative. So, if one uses the IF of the journals where a scholar publishes as a proxy of the impact of his/her research work, one gets a very partial (and frequently unfair) evaluation. As a proxy of the impact of a paper one should use the number of citations collected by the paper, rather than the IF of the journal where the paper was published. This motivated us to define a new measure of individual scholarly impact, the {\it Author Impact Factor} (AIF), which is computed just like the IF, where 
instead of the papers published in a journal one considers the papers of an author. Basically AIF expresses the current impact of papers published by authors in recent years, so it is
a tool to monitor the evolution of the performance of the impact of a scholar's output. The concept of AIF has already appeared before~\cite{MountSinai,jhu,Petersen_Reputation_2013}, 
but we are not aware of any empirical study of its properties, a gap filled by this paper.

In the literature there are several metrics of individual impact. 
Many of these are ranking measures providing quantitative estimates of the relative importance of a scientist~\cite{garfield_citation_1979,radicchi_diffusion_2009}.
The recently introduced $h$-index~\cite{hirsch_index_2005}, which combines the impact of the papers of a scientist with his/her productivity, is by far the most popular. Its fame has led to the proposal of various other related indexes. Such refined measures take into account factors such as the number of co-authors of a publication~\cite{batista_is_2006}, the average number of citations received by a scientist~\cite{egghe_theory_2006}, etc. However most of these refined measures are found to be correlated to the original $h$-index and thus provide little added information~\cite{bornmann_multilevel_2011}. In most cases even a scientist's $h$-index is correlated with the square root of the total number of citations received by the scientist~\cite{hirsch_index_2005}. A common feature of all these metrics is that they are cumulative measures and thus combine all the works done by a scientist during his/her whole research career. However, research productivity and impact vary with time, with different scientists having distinct career trajectories. 

Some successful scientists have made only one major contribution in their career whereas others have been productive throughout their career. The productivity of scientists changes with time. Nobel Laureates often publish fewer papers after they were awarded the prize. This is partly due to other responsibilities and mobilities associated with the prize~\cite{borjas_prizes_2013}. Although prizes are given for past contributions, other decisions such as funding, hiring, etc., are based on current performance. As the correlation between the impact of past and future output is low~\cite{penner_predictability_2013}, measures considering the overall past research performance cannot be taken as a proxy of current or future achievements. Thus there is a big need of dynamic measures providing the current impact of a scholar.  

Since metrics have started to be heavily used in evaluations, scholars have tried to ``adapt'' to the system, making publication choices aiming at the maximization of popular indicators, especially the h-index. As a consequence, the number of papers published every year keeps increasing exponentially~\cite{larsen_rate_2010}.  However, not all of these publications are of high quality. Cumulative measures such as the $h$-index consider only the best publications and do not penalize low quality work. Therefore, many practices such as honorary authorship, publishing immature work and micro publication have been encouraged. This has led to a major proliferation of low quality work, motivated by the hope that even incremental papers have a small chance to attract citations, from which the whole output of the scholar may benefit.

One remedy for these bad practices would be rewarding high quality work and, consequently, penalizing negligible and/or incremental papers. This might refrain scientists from publishing low quality work. Some of the potential breakthrough work might not be published, as the scholar might judge it to be of low quality. However, most of the time a scientist is able to judge the quality of his/her own work, though a prediction of its future impact is difficult. 

The proposed AIF has three major advantages over competitive indicators:
\begin{enumerate}
\item{AIF is a dynamic index, so it can follow the evolution of the impact of a scholar, and state how ``hot'' he/she currently is, if the scholar is in a rising or declining phase and if there is room for improvement. Such considerations might be decisive in hiring decisions, especially for young scientists. Current metrics, like the h-index, are not able to do that.}
\item{AIF averages the number of citations received by all papers published by an author in a given time window, so it is high if those papers are well cited, whereas low quality work would keep the score down. This might incentivize focusing on high quality research.}
\item{AIF is defined just like the IF, so its computation can easily be implemented on most bibliographic portals, like the Web of Knowledge.}
\end{enumerate}

In the next section we shall compute the AIF for Nobel Prize Laureates in various disciplines, and compare it with competitive metrics. 
Then we shall discuss the results and their implications.

\begin{figure}
  \includegraphics[width=0.99\linewidth]{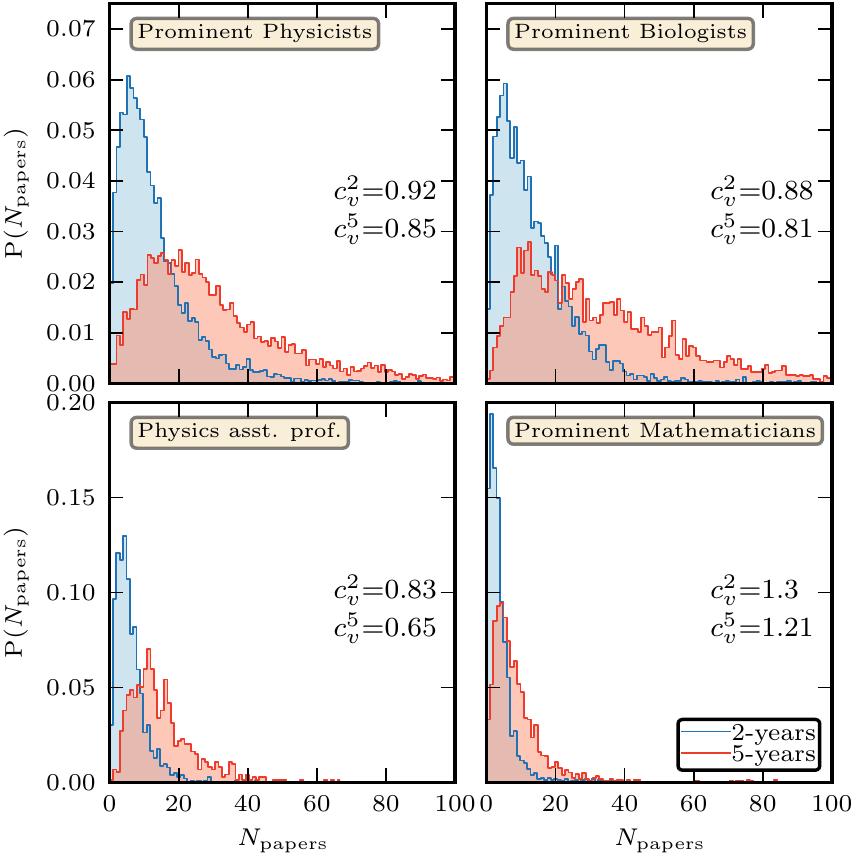}
  \caption{The distribution of the number of papers published by a scientist in 2 years (blue) and 5 years (red) period. The coefficient of variation $c_v$ for each distribution is also indicated.  The values of $c_v$ show that the fluctuations of the number of papers published in a given time window gets smaller when the time window gets large. To avoid the tedious problem of disambiguation of authors' names, we took lists of disambiguated high profile scientists in Physics, Biology and Mathematics that were used in recent works like, e.g. Ref.~\cite{penner_predictability_2013}. A detailed description is provided in the Supplementary Information.}
\label{fig1}
\end{figure}
\begin{figure}
  \includegraphics[width=0.99\linewidth]{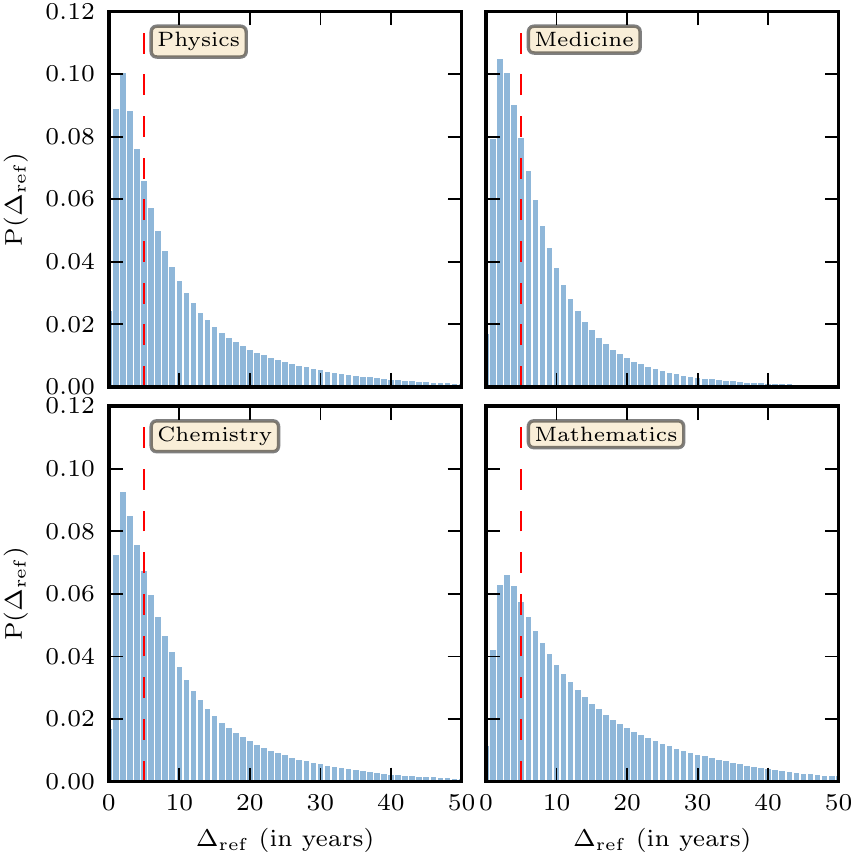}
  \caption{The distribution of the difference between the publication year of the citing article and the publication year of the cited article, for different subjects. This suggests proper aggregation periods to compute the author impact factor. The distribution is generated from papers published between 1980-2010. The red dashed line indicates the 5 years period we choose to compute the AIF: the probability of having laps longer than 5 years is exponentially suppressed.}
\label{fig2}
\end{figure}
\section*{Results}
The AIF of an author $A$ for year $t$ is defined as follows: Let $N^{\Delta t}_{\mathrm{c}}(t)$ be the number of 
times articles published by $A$ in year $[t-\Delta t, t-1]$ were cited during year $t$. And let $N^{\Delta t}_{\mathrm{p}}(t)$ be 
the total number of articles published by the author in $[t-\Delta t, t-1]$. Then, the author impact factor for year $t$ is $N^{\Delta t}_{\mathrm{c}}(t)/N^{\Delta t}_{\mathrm{p}}(t)$.

At variance with the journal impact factor, where one considers all the papers published in a journal in two years, which are typically many, for authors these numbers may be low and there can be considerable fluctuations in the publication rate, especially if one wishes to adopt the measure to authors belonging to different disciplines. In Fig.~\ref{fig1} we show the distribution of the number of papers published in different aggregation periods by high profile scientists of different disciplines.  The fluctuation of the number of papers is indicated by the coefficient of variation $c_v$, i. e. the ratio of the standard deviation $\sigma$ by the mean $\mu$ ($c_v=\sigma/\mu$).  We see that $c_v$ decreases if one aggregates over longer time periods. 
The variation in $c_v$ across the two aggregation time windows is especially notable for the Physics assistant professors, who are relatively young and hence 
the fluctuations are considerably reduced during the 5-year aggregation.

Furthermore, papers usually cite recent publications, so few years after publication the number of cites received by a paper in a year decreases. In Fig.~\ref{fig2} we plot the distribution of the difference in the publication year of the citing article and the cited article. The distribution varies with the discipline. However, we can see that large gaps are suppressed. Hence, the aggregation period should not be very large, as in that case papers published in the initial part of the period would be hardly cited after that. 
During the period 1980-2010 for Physics about 44\% of the citations go to articles published during the immediate past 5 years. 
For Medicine, Chemistry and Mathematics the respective fractions are 47\%, 41\% and 30\%. As these fractions are considerable, we limit our aggregation period to 5 years. 
Another reason for choosing the 5 years time window is to make the AIF congruent to the 5-years journal impact factor provided by Thomson Reuters. 
In the Supplementary Information we also show the AIF computed in a 2 years period (like the 2 year journal impact factor), which shows relatively more fluctuations.

\begin{figure}
  \includegraphics[width=0.99\linewidth]{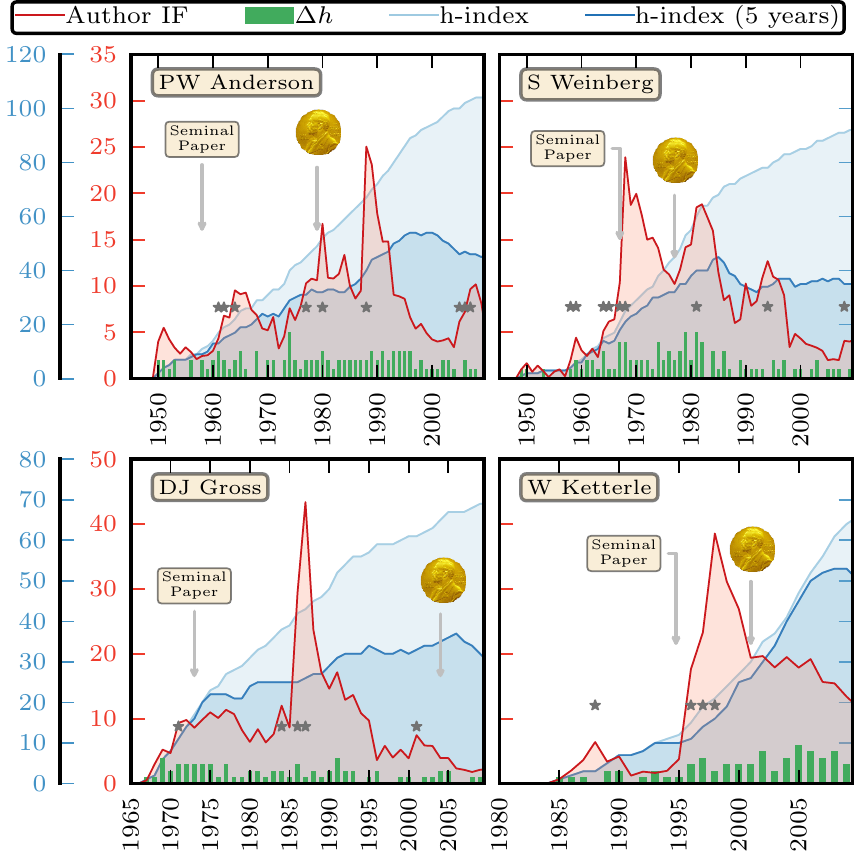}
\caption{Evolution of the author impact factor of four Nobel Laureates in Physics (red curve), compared to the evolution of the $h$-index of the author, 
of its yearly variation $\Delta h$ and of the 5-year $h$ index. The arrows of the diagram indicate significant moments in the career of the scientist,
like the year when he/she received the award and the year when the awarded paper was published. Those years in which the AIF of a scientist significantly 
differs from his/her average AIF of the past 5 years are marked by a star.}
\label{fig3}
\end{figure}

\begin{figure}
  \includegraphics[width=0.99\linewidth]{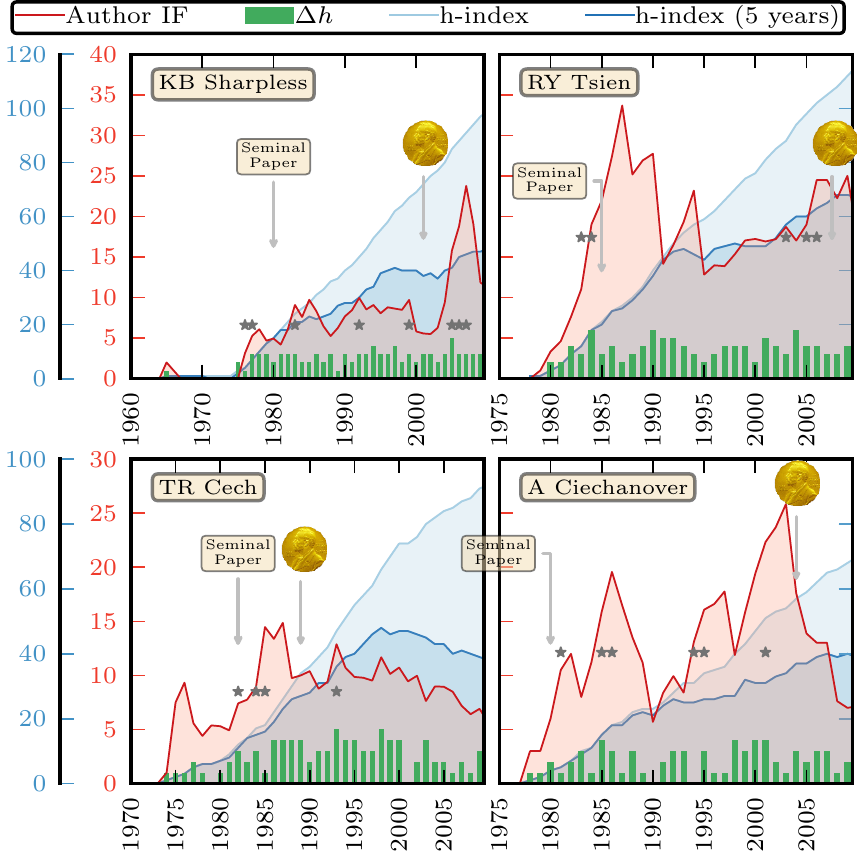}
\caption{Same as Fig. 3, for Nobel Laureates in Chemistry.}
\label{fig4}
\end{figure}

\begin{figure}
  \includegraphics[width=0.99\linewidth]{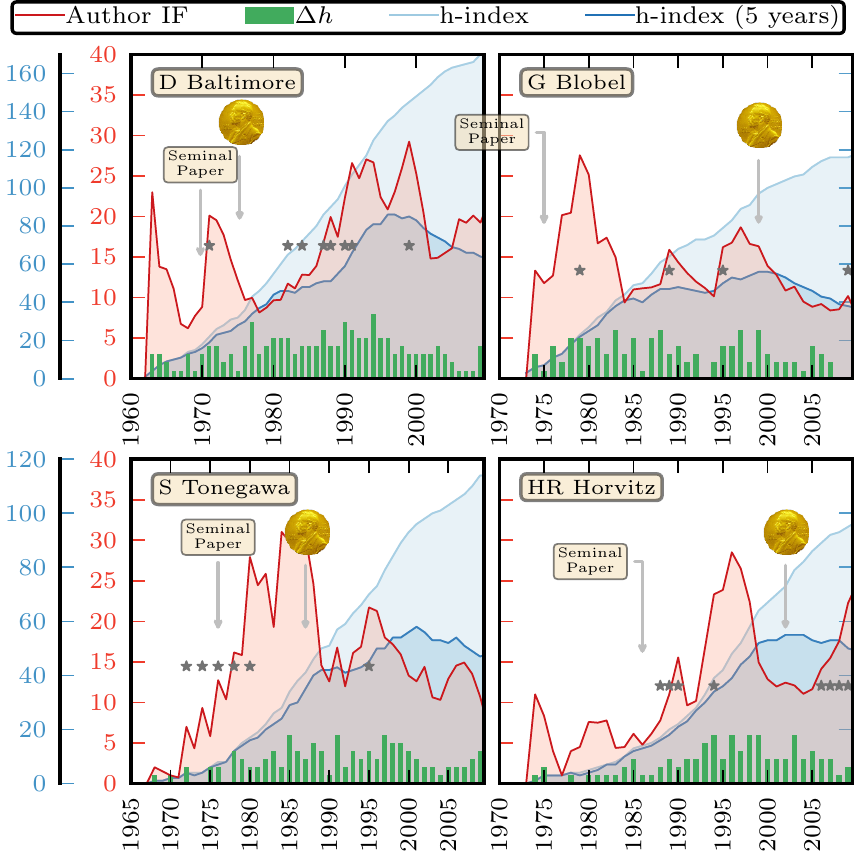}
\caption{Same as Fig. 3, for Nobel Laureates in Physiology or Medicine.}
\label{fig5}
\end{figure}

In Figs.~\ref{fig3}, \ref{fig4} and \ref{fig5} we show the time evolution of the AIF of Nobel Prize Laureates in Physics, Chemistry and Physiology or Medicine.  We report the evolution of other measures as well, for comparison.
First, we consider the $h$-index and its yearly increment $\Delta h$. However, for most scientists the yearly increase in $\Delta h$ fluctuates between $0$ and $3$~\cite{penner_predictability_2013}. Most importantly, neither metric considers the current impact of a scientist, as $\Delta h$ and the corresponding $h$-index could also increase due to papers published a long time back but only cited recently. Even the $h$-index for the last five years is a similar measure. It is defined as the number $h_5$ of papers that have received at least $h_5$ citation in the five years preceding the reference year $t$. These papers could have been published by the scientist at any time during his career and hence $h_5$ does not reveal the true current impact of the scientist in those five years. 

Indeed, we find that the history of the AIF presents a rich structure which reflects the evolution of the careers of the scientists. The other metrics, instead, have a rather plain profile, from which it is difficult to infer anything, except for the moment in which the careers started to fly, possibly. We also marked the years in which the AIF of a scientist significantly differs from his past AIF. 
The past AIF is determined using a simple 5-year moving average (see Methods for detail). 
We now briefly discuss the correlation between the AIF and the careers of the prominent scholars of Figs. 3, 4 and 5.

\subsection*{Nobel Laureates in Physics}
Philip Warren Anderson was awarded the Nobel Prize in 1977 for his investigations of the electronic structure of magnetic and disordered systems.  His pioneering work on the theories of localization is exposed in papers published between  1958 and 1961. Indeed, in the following years the AIF displays a significant bump. However, Anderson has also made fundamental contributions on symmetry breaking, the theory of spin glasses and the scaling theory of localization in the 1970s, which correspond to a major bump around 1980. In the 1980s he also contributed to the field of high-temperature superconductivity, which is reflected by the pronounced spike of the AIF around 1990. On the contrary, the other indicators reported in the plot do not reveal such highlights.

Steven Weinberg is known for his contribution in pion scattering developed in 1966 and electromagnetism and weak force unification theory developed in 1967. The theory was verified in 1973 by the experimental discovery of weak neutral currents and Weinberg was awarded the Nobel Prize in Physics, in 1979. Indeed, the most pronounced spike in Weinberg's AIF profile is around 1970. In 1979 he published his pioneering work on the re-normalization aspects of quantum field theory, which allowed the development of an effective theory of quantum gravity and in 1980 he showed that the graviton cannot be a composite particle in a relativistic quantum field theory (Weinberg-Witten theorem). This corresponds to the second most relevant peak, which is centered around 1980.  

David Jonathan Gross was awarded the 2004 Nobel Prize in Physics, with his former PhD student Frank Anthony Wilczek and Hugh David Politzer, for the discovery of asymptotic freedom in the strong interaction between color charges, which was published in 1973 and was crucial  for the development of quantum chromodynamics. The bump in the early 1970s is fed by the citations to that paper (alongside citations to earlier, less prominent, contributions).  In 1985, he was one of the first scholars to develop heterotic string theory, which fueled the first superstring revolution. The impact of this paper is shown by the sharp peak right after 1985.

Wolfgang Ketterle's research focuses on experiments that trap and cool atoms to temperatures close to absolute zero, and in 1995 he establishes the Bose-Einstein condensation in these systems. He was awarded the Nobel Prize in Physics in 2001 for these achievements. The big peak of the AIF following the publication of the 1995 paper clearly indicates the impact of that work.

\subsection*{Nobel Laureates in Chemistry}
Karl Barry Sharpless is known for his work on asymmetric oxidation (Sharpless epoxidation, Sharpless asymmetric dihydroxylation, Sharpless oxyamination) and shared the Nobel Prize in 2001.  The impact of his seminal work yields the bump in the mid 1980s. In 2001 he published a seminal paper on click chemistry, which is revealed by the prominent spike in the mid 2000s.

Roger Yonchien Tsien was awarded the 2008 Nobel Prize in chemistry for his discovery and development of the green fluorescent protein (GFP).  Tsien's Nobel paper was published in 1985, and its early citations feed the major peak in his AIF profile.  Thomas Robert Cech won the 1989 Nobel Prize in Chemistry for the discovery of the catalytic properties of RNA. 
Cech discovered that RNA could itself cut strands of RNA, which showed that life could have started as RNA. This work, published in 1982, is responsible for the most prominent bump in the AIF evolution.

Aaron Ciechanover won the Nobel Prize in Chemistry in 2004 for characterizing the method that cells use to degrade and recycle proteins using ubiquitin. The ubiquitin-proteasome pathway has a critical role in maintaining the homeostasis of cells and is believed to be involved in the development and progression of diseases such as: cancer, muscular and neurological diseases, immune and inflammatory responses. The seminal paper by Ciechanover was published in 1980, which is followed by the first major bump in the AIF profile of the scientist.  His other important work on the ubiquitin-proteasome pathway and pathogenesis of human diseases, presented in papers published in the 1990s, yields the second large enhancement of the AIF.

\subsection*{Nobel Laureates in Physiology or Medicine}
David Baltimore was awarded the Nobel Prize for Physiology or Medicine in 1975 for his discovery of reverse transcriptase.  The Nobel paper, published in 1970, feeds the second early peak of Baltimore's AIF profile.  His earlier papers on macro-molecular synthesis (1962) and RNA polymerase (1962, 1963) were highly cited too, and are responsible of the first sharp peak. In 1986, he co-discovered NF-$\kappa$ (nuclear factor kappa-light-chain-enhancer of activated B cells), a protein complex that controls the transcription of DNA. This is revealed by the most pronounced bump of the AIF, from the late 1980s.

G\"unter Blobel was awarded the 1999 Nobel Prize in Physiology or Medicine for the discovery of signal peptides. The paper was published in 1975 and is associated to the most important structure of the AIF profile.
Susumu Tonegawa won the Nobel Prize for Physiology or Medicine in 1987 for his discovery of the genetic mechanism that produces antibody diversity, presented in a paper published in 1976. Indeed, the most relevant bump of Tonegawa's AIF profile lies in the years following 1976. 

Howard Robert Horvitz won the 2002 Nobel Prize in Physiology or Medicine for his work on the genetics of organ development and programmed cell death in {\it C. elegans}. In that paper, appeared in 1986, he did the first characterization of genes regulating cell death, ced-3 and ced-4, in the nematode worm.  In fact, the period following 1986 hosts the second highest peak of Horvitz's curve.  A seminal 1993 work on amyotrophic lateral sclerosis is responsible for the most pronounced peak (around 1995). Finally, the rising trend of the latest years follows recent important work on Micro-RNA expression profiles classifying human cancers.


\section*{Discussion}
We introduced Author Impact Factor (AIF), a dynamic index to quantify the impact of recent work of scientists, enabling one to track the evolution of the performance
of scholars along their careers, especially trends. 
We are aware that the literature on scientometrics is full of performance indicators, and we are generally against an indiscriminate proliferation of metrics.
However, AIF fills an important gap, as current indicators of individual performance are not able to follow the dynamics of careers, and 
are not sufficiently sensitive to major events, like sharp variations in the citation flows to an author's work, e.g. following the publication of groundbreaking papers.
We have given striking evidence of this in our examination of the careers of Nobel Laureates.

AIF is the analogue for authors of what the impact factor is for journals. The journal impact factor, despite its many limitations, 
is a very familiar concept to scientists, and regularly used by academic and research institutions in performance evaluations.
The simplicity of the AIF can favor its adoption, especially by bibliographic portals, which could easily compute the measure, 
as they currently compute other metrics, like the $h$-index.

In addition, the AIF has the following benefits:
\begin{enumerate}
\item{It allows to check whether major papers are just lucky accidents in otherwise modest careers, or the highlights of 
a consistent production of significant work. This of course could also be revealed by checking the number of citations of each paper of the author.}
\item{It penalizes low quality work, so it might help to contrast the proliferation of negligible papers we have been witnessing in the last years,
prioritizing high quality science. Likewise, since the focus would be on the quality and not on the number of papers, practices like 
honorary authorship would be discouraged.}
\item{The criteria typically adopted by academic/research institutions and funding organizations are dominated by cumulative metrics, that consider the 
whole scientific career. Although this is not a bad thing in itself, this feature has certain drawbacks. Cumulative measures 
increase with the career age of a scientist. Thus they are biased towards senior scientists and would penalize junior scientists, even when the senior 
fellow's performance is declining whereas the junior's performance is skyrocketing.
The AIF would show, among any two scientists, who is (has been) `` hotter'' in the latest years (hence a more promising recruit or grant holder), regardless of career age.}
\end{enumerate}

Here, we use a time window of 5 years to calculate the AIF. Although for prominent scientists with a large number of yearly publications, a shorter time window yields similar results. However, young scientists with relatively fewer yearly publications show more fluctuations if the AIF is calculated using a shorter 2-year time window. Further, the 5-year time window is also used by Thomson Reuters to calculate the 5-year Journal Impact Factor. One possible drawback of AIF is that it only captures the citations received by a paper within 5 years from its publication. Thus, it undermines those publications
receiving the bulk of their citations after this time window (``sleeping beauties''). One way to include these publications is to consider the immediacy of a paper, i.e., the time for a paper to reach its citation peak, and count the number of citations up to the peak~\cite{wang_quantifying_2013}. Although this allows to define a time-window-free AIF, it needs the citation history of each publication over a much longer time window. Further, the simplicity of AIF is lost as the AIF of a given year is not fixed and may change later due to the citation boost to a paper occurring at much later time period.
An alternative way to estimate the dynamic productivity of a scientist is to measure the time period in which he or she has published papers responsible for half of the total number of citations~\cite{popov_a_2005}. Although this provides a way to distinguish scientists with long steady careers from those with limited career activity, it captures neither the actual evolution of their career nor how productive they were during this career period.

Other related measures could be the incremental $h$-index, where the $h$-index is calculated for papers published in the time window $[t-\Delta t, t]$ 
and one counts only the citations received during this period. This is different from $\Delta h$, that we have seen above, as in that case the $h$-index can also increase due to papers published before the reference period. However, measures related to the $h$-index are penalized by the fact that the $h$-index is integer, and its increments are typically low numbers, so they have little discriminative power. One can also consider the average number of citations of papers published in year $t$ within $\Delta t$ years of their publication. However, as the same paper and its citations are considered in two consecutive time windows, there would be some inherent correlations. Further, the citation rate depends on the author's reputation and hence it generally increases with his or her scientific age~\cite{Petersen_Reputation_2013}.  Other metrics are i10, the number of publications that received at least 10 citations in the last $\Delta t$ years and the total number of citations in the last $\Delta t$ years. The threshold of 10 citations is however arbitrary, and neither measure penalizes low quality work. Thus, the AIF provides a unique dynamic metric that goes beyond the existing cumulative measures and their dynamic extensions.

\section*{Methods}
\textit{Dataset Description.} We analyzed the careers of 12 Nobel Laureates, 4 each in Physics, Chemistry and Physiology or Medicine. In addition, we also analyzed the publication profiles of 350 scientists divided into 4 broad categories: 100 prominent physicists, 100 prominent cell biologists, 50 prominent pure mathematicians and 100 assistant professors in physics. 
The first three categories comprise the top-cited scientists in their respective fields. For each author we extract all their publications included in the database of Thomson Reuters (TR) Web of Knowledge historical publication and conference proceedings database. For each of these papers we extract its year of publication and the corresponding citations to that publication. For author impact factor only ``Citable items'' (articles, reviews, proceedings, or notes; not editorials and/or letters to editor) were considered. For more information on author selection and disambiguation method, see the Supplementary Material.  

We also analyzed all publications (articles, reviews and editorial comments) from 1980 till the end of 2010 included in the database of the TR. For each publication we extract its year of publication, list of references and the subject category of the journal in which it is published. We parsed the references of each publication and determined their year of publication. Based on the subject category of the journal of the publication, the papers were categorized in Physics, Medicine, Chemistry and Mathematics.

\textit{Significant Peaks.}
To determine whether the AIF of a scientist in year $t$ significantly differ from his/her immediate past we use the following procedure: The past 5-year AIF is determined using a simple moving average (SMA) and used as a proxy of the immediate past impact. The SMA$(t)$ is the unweighted mean of the previous 5 AIF's, i.e., SMA$(t)$ = [AIF$(t-1) + \dots + $AIF$(t-5)$]/5. We also determine the standard deviation $\sigma(t)$ in the AIF during the period $[t-1,\dots,t-5]$. Then, the $z$-score for a year $t$ is given by $z(t)=\frac{\mathrm{AIF}(t)-\mathrm{SMA}(t)}{\sigma(t)}$. If the $z$-score of a year $t$ exceeds 2.326, corresponding to a chance lower than 1\% that the value of the AIF$(t)$ occurred randomly, we conclude that the AIF of year $t$ is significantly different from the past years.

\section*{Acknowledgments}
We thank A.M. Petersen for providing some of the data used in this research.
Some data included herein are derived from the Science Citation Index Expanded, Social Science Citation Index and Arts \& Humanities Citation Index, prepared by Thomson Reuters, Philadelphia, Pennsylvania, USA, Copyright Thomson Reuters, 2011. We gratefully acknowledge KNOWeSCAPE, COST Action TD1210 of the European Commission, for fostering interactions with leading experts 
in science of science, which inspired this work.

\bibliography{nobelPapers,sc}

\appendix

\begin{figure}
  \includegraphics[width=0.99\linewidth]{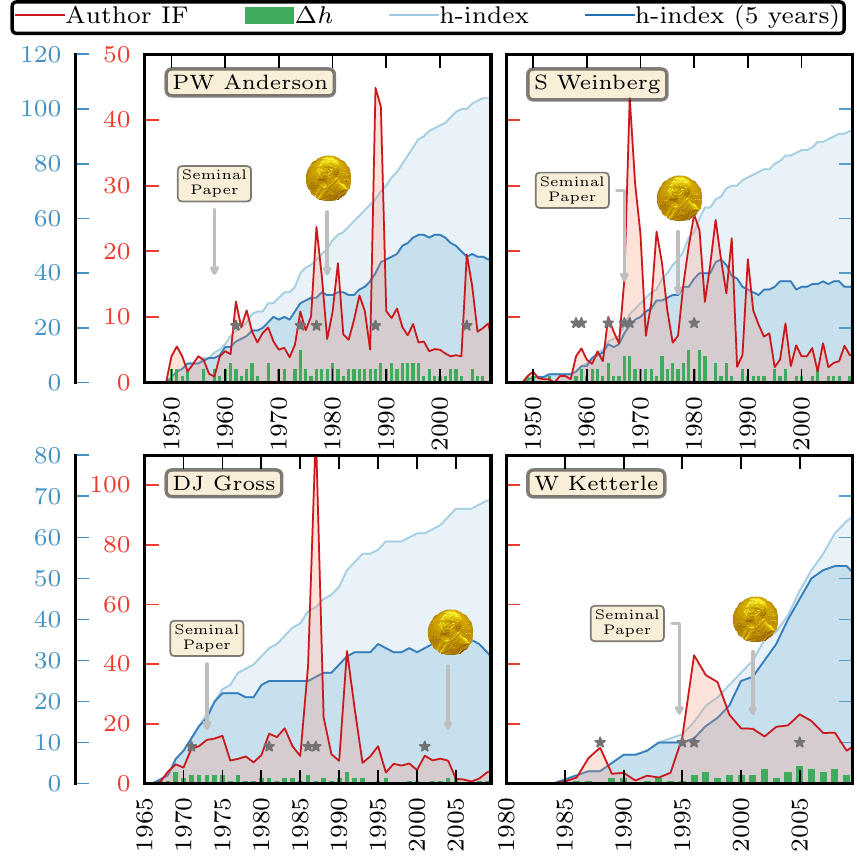}
\caption{Evolution of the author impact factor of four Nobel Laureates
  in Physics (red curve), compared to the evolution of the $h$-index
  of the author, of its yearly variation $\Delta h$ and of the 5-years
  $h$ index. In this case the aggregation period $\Delta t$ of 2 years
  was used to calculate the author impact factor. Those years in which the AIF of a scientist significantly 
differs from his/her average AIF of the past 5 years are marked by a star.}
\label{figS1}
\end{figure}
\begin{figure}
  \includegraphics[width=0.99\linewidth]{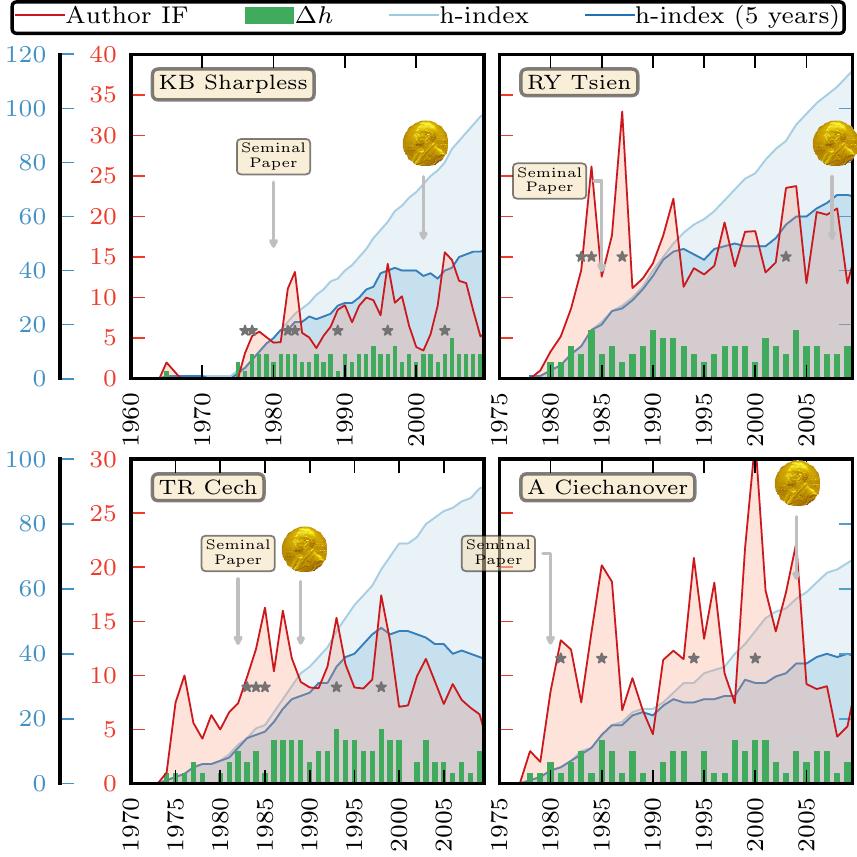}
\caption{Same as Fig. 3, for Nobel Laureates in Chemistry.}
\label{figS2}
\end{figure}

\begin{figure}
  \includegraphics[width=0.99\linewidth]{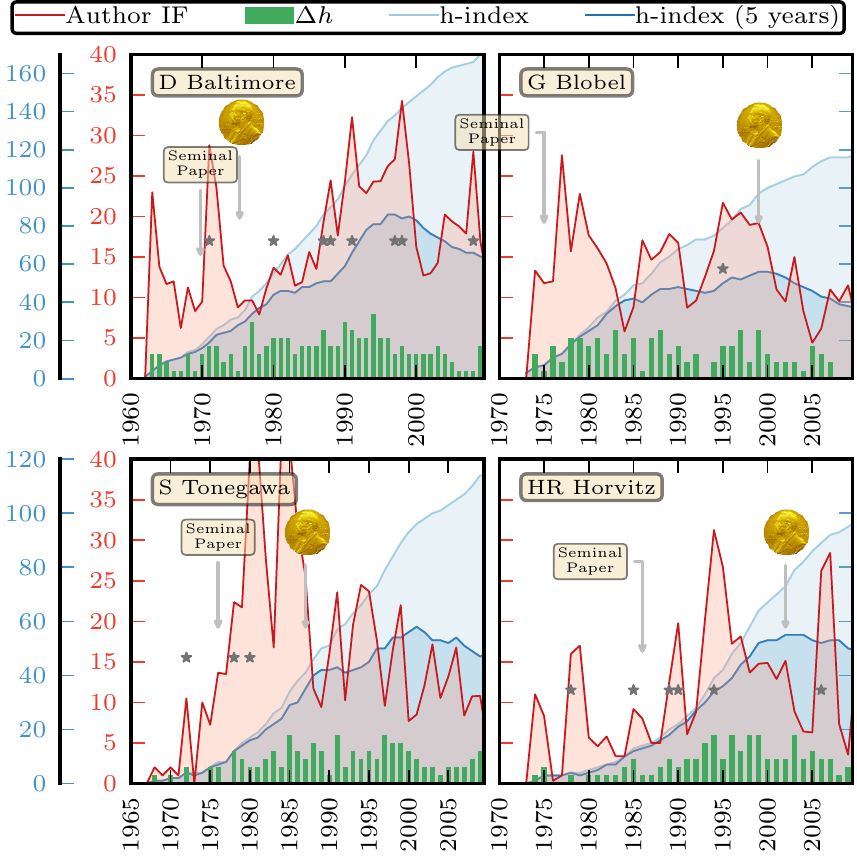}
\caption{Same as Fig. 3, for Nobel Laureates in Physiology or Medicine.}
\label{figS3}
\end{figure}

\section{Methods}
Here we use disambiguated  ``distinct author'' data from Thomson
Reuters (TR) Web of Knowledge, {\tt isiknowledge.com} using 
their matching algorithms to identify publication profiles of distinct
authors. Further, we use its portal {\tt ResearcherID.com}, 
where users upload and maintain their publication profiles. We
consider a total of 12 Nobel Laureates, 4 each in Physics, Chemistry
and 
Physiology or Medicine. In addition we analyze 550 scientists
divided into 4 categories. For the selection of high-impact
physicists, 
we consider the 100 most prolific authors based on their publications in {\it
  Physical Review Letters} over the 50-year period 1958-2008.
For the selection of high-impact cell biologists we choose the 100
most prolific scientists based on publications in the journal {\it CELL}. For the
selection of 
high-impact mathematicians we selected the 50 authors with the highest
number of
publications in the prestigious journal {\it Annals of Mathematics}. 
We also consider 100 relatively young assistant professors from
physics. To select the scientists in this dataset, we choose two
assistant 
professors from each of the top 50 U.S. physics and astronomy departments ranked according to the magazine {\it U.S. News}. 

To categorize each paper according to its field of publication
we use the TR subject categories. We then aggregated these subject 
categories into broader scientific fields. A detailed description is provided in Table~\ref{tab:cat}

\section{Results}
We used the aggregation period of 2 years to calculate the author
impact factor (AIF) for the Nobel laureates in Physics (Fig.~\ref{figS1}), 
Chemistry (Fig.~\ref{figS2}), and Physiology or Medicine
(Fig.~\ref{figS3}). Although there are relatively more fluctuations, 
the variation of the AIF(2 years) is qualitatively similar to the variation of AIF(5 years). 
The 2-year window is however short for most scientists, especially for the young ones. As the paper only shows the AIF for the most prominent scientists, both the 2-year and 5-year AIF yield similar results.


\begin{longtable*}{l|>{\footnotesize}l}
\toprule
Fields & TR subject categories \\ 
\colrule
\multirow{19}{*}{Physics} & IMAGING SCIENCE \& PHOTOGRAPHIC TECHNOLOGY\\ 
 & PHYSICS, APPLIED\\
 & OPTICS\\
 & INSTRUMENTS \& INSTRUMENTATION\\
 & PHYSICS, CONDENSED MATTER\\
 & PHYSICS, FLUIDS \& PLASMAS\\
 & PHOTOGRAPHIC TECHNOLOGY\\
 & PHYSICS, ATOMIC, MOLECULAR \& CHEMICAL\\
 & ACOUSTICS\\
 & PHYSICS\\
 & PHYSICS, MATHEMATICAL\\
 & MECHANICS\\
 & PHYSICS, NUCLEAR\\
 & SPECTROSCOPY\\
 & THERMODYNAMICS\\
 & PHYSICS, PARTICLES \& FIELDS\\
 & NUCLEAR SCIENCE \& TECHNOLOGY\\
 & PHYSICS, MULTIDISCIPLINARY\\
 & ASTRONOMY \& ASTROPHYSICS\\
\colrule
\multirow{6}{*}{Mathematics} & STATISTICS \& PROBABILITY\\
 & MATHEMATICS, APPLIED\\
 & MATHEMATICS, INTERDISCIPLINARY APPLICATIONS\\
 & LOGIC\\
 & MATHEMATICS\\
 & MATHEMATICS, MISCELLANEOUS\\
\colrule
\multirow{10}{*}{Chemistry} & CHEMISTRY, INORGANIC \& NUCLEAR\\
 & ELECTROCHEMISTRY\\
 & CHEMISTRY, PHYSICAL\\
 & CHEMISTRY, ANALYTICAL\\
 & POLYMER SCIENCE\\
 & CHEMISTRY, MULTIDISCIPLINARY\\
 & CRYSTALLOGRAPHY\\
 & CHEMISTRY, APPLIED\\
 & CHEMISTRY\\
 & CHEMISTRY, ORGANIC\\
\colrule
 & CYTOLOGY \& HISTOLOGY\\
 & BIOCHEMISTRY \& MOLECULAR BIOLOGY\\
 & CELL BIOLOGY\\
 & BIOCHEMICAL RESEARCH METHODS\\
 & CELL \& TISSUE ENGINEERING\\
 & MATHEMATICAL \& COMPUTATIONAL BIOLOGY\\
 & BIOPHYSICS\\
 & BIOMETHODS\\
 & MICROSCOPY\\
 & ENGINEERING, BIOMEDICAL\\
 & IMMUNOLOGY\\
 & MEDICAL LABORATORY TECHNOLOGY\\
 & MEDICINE, RESEARCH \& EXPERIMENTAL\\
 & PARASITOLOGY\\
 & PHYSIOLOGY\\
 & ANATOMY \& MORPHOLOGY\\
 & PATHOLOGY\\
 & ONCOLOGY\\
 & RHEUMATOLOGY\\
 & VASCULAR DISEASES\\
 & PSYCHIATRY\\
 & GERIATRICS \& GERONTOLOGY\\
 & DENTISTRY, ORAL SURGERY \& MEDICINE\\
 & OPHTHALMOLOGY\\
 & DENTISTRY/ORAL SURGERY \& MEDICINE\\
 & MEDICINE, LEGAL\\
 & EMERGENCY MEDICINE \& CRITICAL CARE\\
 & CLINICAL NEUROLOGY\\
 & TRANSPLANTATION\\
\multirow{31}{*}{Physiology or Medicine} & HEMATOLOGY\\
 & INFECTIOUS DISEASES\\
 & RESPIRATORY SYSTEM\\
 & PERIPHERAL VASCULAR DISEASE\\
 & MEDICINE, GENERAL \& INTERNAL\\
 & PEDIATRICS\\
 & EMERGENCY MEDICINE\\
 & INTEGRATIVE \& COMPLEMENTARY MEDICINE\\
 & GASTROENTEROLOGY \& HEPATOLOGY\\
 & DERMATOLOGY\\
 & REHABILITATION\\
 & ANESTHESIOLOGY\\
 & TROPICAL MEDICINE\\
 & MEDICINE, MISCELLANEOUS\\
 & ENDOCRINOLOGY \& METABOLISM\\
 & NEUROIMAGING\\
 & ANDROLOGY\\
 & ORTHOPEDICS\\
 & OBSTETRICS \& GYNECOLOGY\\
 & ALLERGY\\
 & CRITICAL CARE MEDICINE\\
 & OTORHINOLARYNGOLOGY\\
 & RADIOLOGY, NUCLEAR MEDICINE \& MEDICAL IMAGING\\
 & SURGERY\\
 & CARDIAC \& CARDIOVASCULAR SYSTEMS\\
 & DERMATOLOGY \& VENEREAL DISEASES\\
 & AUDIOLOGY \& SPEECH-LANGUAGE PATHOLOGY\\
 & RADIOLOGY \& NUCLEAR MEDICINE\\
 & UROLOGY \& NEPHROLOGY\\
 & CRITICAL CARE\\
 & CARDIOVASCULAR SYSTEM\\
\botrule
\caption{Aggregation of TR subject categories in broader fields.}
\label{tab:cat}
\end{longtable*}

\end{document}